\documentclass[twocolumn,showpacs,preprintnumbers,amsmath,amssymb]{revtex4}

\usepackage{graphicx}
\usepackage{dcolumn}
\usepackage{bm}

\begin{document}


\title{Graphene on the carbon face of SiC: electronic structure modification by hydrogen intercalation}

\author{F. Hiebel, P. Mallet, J.-Y. Veuillen and L. Magaud}
\affiliation{%
Institut N\'{e}el, CNRS-UJF, BP 166, 38042 Grenoble Cedex 9, France
}

\date{\today}

\begin{abstract}
It has been shown that the first C layer on the SiC$(000\overline{1})(2\times2)_C$ surface already exhibits graphene-like electronic structure, with linear $\pi$ bands near the Dirac point. Indeed, the $(2\times2)_C$ reconstruction, with a Si adatom and C restatom structure, efficiently passivates the SiC$(000\overline{1})$ surface thanks to an adatom/restatom charge transfer mechanism. Here, we study the effects of interface modifications on the graphene layer using density functional theory calculations. The modifications we consider are inspired from native interface defects observed by scanning tunneling microscopy. One H atom per $4\times4$ SiC cell ($5\times5$ graphene cell) is introduced in order to saturate a restatom dangling bond and hinder the adatom/restatom charge transfer. As a consequence, the graphene layer is doped with electrons from the substrate and the interaction with the adatom states slightly increases. Native interface defects are therefore likely to play an important role in the doping mechanism on the C terminated SiC substrates. We also conclude that an efficient passivation of the C face of SiC by H requires a complete removal of the reconstruction. Otherwise, at variance with the Si terminated SiC substrates, the presence of H at the interface would increase the graphene/substrate interaction.

\end{abstract}

\maketitle

\section{\label{Introduction}Introduction}

Graphene, a single layer of graphite, has two C atoms per unit cell, usually called A and B, that form a honeycomb lattice. Its low-energy electronic structure exhibits two $\pi$ bands, the $\pi$ band is filled and the $\pi$* band is empty. They touch at the Dirac point, at the corner of the Brillouin zone. The corresponding energy is called the Dirac Energy ($E_D$). Importantly, the dispersion is linear within $\pm0.5$eV with respect to $E_D$\cite{PhysRev.71.622}. The experimental discovery of graphene's unique electronic structure\cite{novoselov-2005-438, zhang-2005-438} and high electronic mobility\cite{suspended} announces a whole new era for physics as well as nanoelectronics\cite{rise, Geim06192009}. Among the fabrication techniques like mechanical exfoliation from a graphite crystal\cite{citeulike:2548167} or catalytic decomposition of hydrocarbons on transition metals\cite{Wintterlin20091841}, graphitization of the polar faces of SiC has the advantage of providing large scale graphene sheets, directly on an insulating substrate\cite{CB}. However, since graphene is an ultimately thin crystal, its environment has to be carefully considered. Proving that point, the highest mobility so far has been achieved for suspended graphene\cite{suspended}.
 Hence, for supported graphene, it is of particular importance to investigate the interface structure and to study how the substrate impacts the electronic properties of the graphene layer.

For graphene grown on the Si terminated substrates (SiC(0001)), the interface is now well characterized. The first carbon layer, possibly with defects\cite{PhysRevLett.105.085502}, strongly interacts with the substrate and forms covalent bonds. This layer yet acts as a buffer layer, so that the second graphitic plane exhibits graphene low-energy electronic properties\cite{mallet:041403, scatteringRutter, brihuega:206802, varchonprl, ripples, mattausch:076802, kimPRL}. Thus, an overall linear dispersion is observed by angle resolved photoemission spectroscopy (ARPES), with a doping in electrons that shifts the Dirac point 0.4 eV below the Fermi energy\cite{BostwickdepotK, Zhou}. 

Intentionally adsorbed species on the top of graphene can then be used to tailor its electronic properties.
By depositing molecules\cite{molec1, molec2}, transition metals\cite{depotTM} or other elements\cite{BostwickdepotK, depotBiAuSb}, one can control the position of the Fermi level without altering the typical band structure of graphene. Adsorbates however are likely to act as scattering centers and increase the elastic scattering rate and then affect the transport properties\cite{ChenNatPhys, depotTM, PhysRevB.81.115453}.
For this reason, modifications of the interface seem more adapted since they occur further from the graphene layer. For example, gold can be introduced at the interface in order to dope the graphene layer with holes\cite{depotBiAuSb}. Recently, passivation of the substrate has been achieved using hydrogen. In this case, the buffer layer is decoupled from the substrate and recovers a low-energy electronic structure similar to pristine (neutral) graphene\cite{expeHintercale, DFTHintercale}.

On the C terminated face (SiC$(000\overline{1})$), the situation is different. For ultra-high vacuum grown samples, coexisting native surface reconstructions of SiC$(000\overline{1})$, namely the SiC($3\times3$)\cite{Hoster3x3} and the SiC$(2\times2)_C$\cite{Seubert}, are found at the interface\cite{hiebelprb1, revstarke}. The graphene-substrate interaction is much weaker than on the Si face. Graphene is almost free-standing on the SiC$(3\times3)$ reconstruction\cite{hiebelprb2} and weakly interacting with the SiC$(2\times2)_C$ reconstruction\cite{hiebelprb1, magaudprb}. Monolayer samples show a band structure that resembles that of free-standing graphene, but with an intrinsic doping in electrons\cite{emtsevprb}. The Dirac point is found 0.2 eV below the Fermi level\cite{CB, emtsevprb}. For this system, effects of modifications of the interface -either intentional or due to native defects- remain to be investigated.

Here we focus on the SiC($2\times2)_C$ interface, which has a simple Si adatom and C restatom structure \cite{Seubert}. 
The electronic properties of graphene on SiC($2\times2)_C$ have already been investigated using ab initio calculations\cite{magaudprb}. For this system, without defects, we have shown that i) the total energy of the system is not sensitive to graphene versus SiC lattice translations, ii) the graphene layer lies far from the substrate ($3.1$\AA), iii)linear dispersion is preserved in the vicinity of the Dirac point, iv) interaction with the adatom states occurs and results in a band anticrossing. It leads to band structure modifications for energies higher than 0.5 eV with respect to $E_D$ and v) there is no charge transfer from the substrate. This latter point is in apparent contradiction with experiments. Indeed, as mentioned above, transport measurements\cite{CB} and ARPES measurements\cite{emtsevprb} indicate a doping in electrons. In particular in ref. \onlinecite{emtsevprb}, for a submonolayer sample with both SiC($3\times3$) and SiC($2\times2)_C$ types of interface, authors find the graphene Fermi level $0.2$ eV above $E_D$. 

In this paper, we show by means of ab initio calculations that a class of interface defects- related to the specific atomic structure of the interface- allows to change the electron doping level of graphene on SiC($2\times2)_C$ ($G\_2\times2$ in the following). These defects could be intentionally generated by chemical modifications but scanning tunneling microscopy (STM) data indicate that they are natively present on as-grown samples. Incidentally, defects observed by STM could explain the discrepancy between the calculated electronic structure for a perfect interface structure and the experimental results mentioned above.

Based on STM data, we propose in section II. C a structural model for the defects in which an adatom dangling bond is saturated by a H atom. 
We then address the effect of these defects on the graphene-substrate interaction, and on the electronic properties of the graphene overlayer by means of ab initio calculations. Finally, we compare ab initio results to STM data in order to check the validity of the defect model.
\section{\label{param}experiment and Calculation details}
\subsection{Experimental aspects}

Substrates are n doped 6H-SiC$(000\overline{1})$ (C face) polished by NovaSiC. Samples are prepared under ultrahigh vacuum (UHV), following a procedure presented in ref. \onlinecite{hiebelprb1} and \onlinecite{hiebelprb2}. Samples are first annealed at $850 ^\circ$C under Si flux in order to get a clean surface. Further annealing at $950-1000^\circ$ C provides a well ordered SiC($3\times3$). Finally, increasing the temperature to $1100^\circ$ C leads to a graphene coverage of a fraction of a monolayer. The growth is controlled by low-energy electron diffraction (LEED) and Auger spectroscopy. For graphitized samples, the LEED pattern shows two surface reconstructions, the SiC($3\times3$)\cite{Hoster3x3} and the SiC$(2\times2)_C$\cite{Seubert}. Already for a submonolayer coverage, the graphene signal is ring-shaped with modulated intensity, indicating a strong rotational disorder within the graphene film\cite{hiebelprb1, emtsevprb, revstarke, hiebelprb2}.

Scanning tunneling microscopy (STM) measurements were performed in situ, under UHV at room temperature. Tips are mechanically cut PtIr. STM gives access to the local density of states (LDOS) at the surface of the samples, with atomic resolution. In the case of graphene on SiC, it is well known that low-bias images give access to the LDOS of the graphene layer while at high bias, graphene becomes transparent and images show the LDOS of the interface, i.e. under the graphene layer\cite{mallet:041403, rutter:235416}. As presented in previous papers\cite{hiebelprb1, hiebelprb2}, the samples show bare SiC($3\times3$) domains and graphene monolayer islands on the SiC($3\times3$) and the SiC$(2\times2)_C$ reconstructions. In agreement with LEED, there is a strong rotational disorder between the graphene islands which gives rise to various moir\'{e} patterns\cite{hiebelprb2}.
\subsection{calculational framework}
Calculations are carried out using the VASP code\cite{vasp}, which is based on the density-functional theory (DFT). We use the generalized gradient approximation\cite{gga} together with ultrasoft pseudopotentials\cite{pseudopot}. The 4H-SiC substrate is modeled by a slab containing 4 SiC bilayers. The backside of the slab is passivated by hydrogen (see Fig. \ref{fig2} (b)). An empty space of 8 \AA\ separates the graphene layer from the next SiC slab. We use a plane wave basis cutoff of 211 eV. The ultrasoft pseudopotentials have been extensively tested\cite{magaudprb, ripples}. Integration over the Brillouin zone is carried out within the Monckhorst-Pack scheme, using a $6\times6\times5$ grid. The K point at the corner of the Brillouin zone is included in the grid in order to get an accurate description of the band structure at the Dirac point. All the structures were fully converged, with residual forces smaller than $0.015$ eV/\AA.

STM images are simulated by cross sections of $|\Psi|^2$ integrated over different ranges of energy, depending on the imaging bias.
\subsection{Modeling of the defects}
\begin{figure}[!h]
\includegraphics[width=0.5\textwidth]{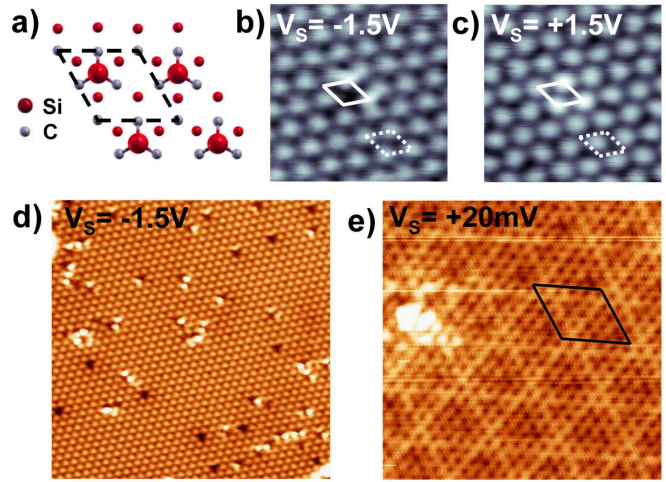}
 \caption{\label{fig1} (Color online) STM images of G\_$2\times2$ islands. Sample biases ($V_S$) are indicated on the images. (a) The SiC($2\times2)_C$ reconstruction. The unit cell is shown in dashed lines. The size of the atoms refers to their height in the cell. Only the topmost SiC bilayer and the Si adatoms are represented. (b) and (c) $4\times4$ nm$^2$ dual high-bias STM images. The SiC($2\times2)_C$ appears. The undisturbed unit cell is given in dashed lines. A typical interface defect is shown at the center, highlighted by the full line cell. (d) $20\times20$ nm$^2$ high-bias STM image. Point defects are seen at the interface. (e) $5\times5$ nm$^2$ low-bias STM image of a G\_$2\times2$ island with no rotation with respect to the substrate. A defect is seen on the left side of the image. Away from the defect, at the atomic scale, the graphene honeycomb lattice is seen. At a larger scale, a periodic pattern ($5\times5$ with respect to the graphene lattice) is visible. Its pseudo-cell is shown in full line. }
\end{figure}

Fig. \ref{fig1} (a) shows the ideal structure of the SiC$(2\times2)_C$ reconstruction. A Si adatom sits in a three-fold coordinated hollow site (H3) and saturates 3 out of 4 C dangling bonds (DB) of the ideal surface\cite{Seubert}. This leads to 2 DB per unit cell, one on the Si adatom and one on the fourth C atom (located at the corners of the diamond cell in Fig. 1 (a)). We call this fourth atom a restatom, by analogy with the dimer adatom stacking fault (DAS) model that describes the Si(111)($7\times7$)\cite{7x7} surface reconstruction, the SiC$(2\times2)_C$ showing however a much simpler structure. From ab initio calculations\cite{magaudprb} and STM measurements \cite{hiebelprb1, PSS}, it is established that a charge transfer occurs, resulting in an empty DB for the Si adatom and a filled DB for the C restatom. Such a charge transfer mechanism is also observed on the Si(111)($7\times7$) and Ge(111)c($2\times8$) surfaces, which involve restatoms and adatoms\cite{Kubby}. Fig. \ref{fig1} (b) and (c) are dual bias STM images that illustrate this statement. In Fig. \ref{fig1} (b), a filled-state image, C restatoms appear at the corners of the dashed-line unit cell. In Fig. \ref{fig1} (c), an empty-state image, the Si adatom appears within the dashed-line unit cell. 

The real interface however exhibits a significant amount of defects. On Fig. \ref{fig1} (d), a high-bias filled-state STM image of G$\_2\times2$, the SiC$(2\times2)_C$ restatom lattice is interrupted by defects that look like ``missing'' restatoms surrounded by  bright spots. 
These defects are still detected on low-bias images, together with the graphene lattice. In Fig. \ref{fig1} (e), the typical low-bias feature of a defect is seen on the left side of the image. Thus, the defect contributes to the DOS at the Fermi level ($E_F$), within the substrate surface bandgap, and could therefore be responsible for the doping of the graphene layer. Note that the graphene island in Fig. \ref{fig1} (e) presents the same geometry as in the calculations (presented below), i.e. the graphene and the SiC surface lattices are aligned. Away from the defect, the graphene low-energy LDOS appears as a honeycomb lattice, modulated by a moir\'{e} pattern which corresponds to the graphene $5\times5$ /SiC $4\times4$ common cell\cite{PSS}. Its pseudo-cell is shown on the image.

Although STM does not allow chemical characterization of the surface, it can provide useful hints on the structure of the defects in the real interface. 
Fig. \ref{fig1} (b) and (c), show a typical defect, within the solid-line diamond cell.
On filled-state STM images (Fig. \ref{fig1} (b)), the defects generally appear as a ``missing'' restatom, with one or more of its nearest adatom neighbors visible at \itshape both \upshape sample bias polarity (at the corners of the solid-line diamond cell). 
On empty-state STM images(see Fig. \ref{fig1} (c)), we see the regular lattice of adatoms, with stronger signal on the 3 neighbors of the defect. These empty-state and filled-state features can arise from an impurity that saturates the restatom DB, and thus hinders the charge transfer from the neighboring adatoms. We propose a simple model to feature these defects that involves a H atom, which is also interesting since for the Si face of SiC, hydrogen decouples graphene from its substrate or saturates interface DB \cite{guisingerH}. Note that such a cancellation of charge transfer has been observed on the Ge(111)c$(2\times8)$ surface after hydrogen adsorption\cite{dujardinPRB63}. In this case, the presence of the defect leads to a similar empty/filled states contrast on restatoms/adatoms, which further supports our model. 

In order to build the calculation cell, we start from the converged G$\_2\times2$ structure presented in ref \cite{magaudprb}, namely a 5x5 graphene cell on top of a 4x4 SiC supercell, without rotation (see Fig. \ref{fig2} (a)). The mismatch between the two supercells is smaller than $0.2$ \%. We then introduce some defects at the interface.  One H atom is introduced in the G$\_2\times2$ structure, on top of the restatom, midway between the restatom and the graphene layer i.e. at 2 \AA\ from the restatom to be saturated. Since the supercell contains 4 SiC$(2\times2)_C$ cells, there are 4 available locations for the defect - labeled C1, C2, C3 and C4 in Fig. \ref{fig2} (a). The 3 configurations C1, C3 and C4 appear similar with a defect located under a C-C bond. The defect in C2 is located under a graphene hexagon. All 4 configurations were studied and actually lead to similar electronic properties. We therefore present results only for the C1 configuration, some results for the C2 configuration are given in the supplementary information\cite{EPAPS}.

\section{\label{results} Results}
\subsection{\label{struct_relax}Atomic structure}
After relaxation, the H atom has indeed formed a bond with the restatom and is now located only 1.10 \AA\ above it. Some other modifications of the atomic positions are also noticeable, but they do not significantly depend on the defect location.
The mean graphene layer-adatom distance (initially of 3.10 \AA) dropped by 0.15 \AA. On the lower half of the 5x5 graphene supercell, atoms of the graphene layer have gone down toward the substrate, slightly increasing the already existing layer corrugation, from $0.15$ \AA\ peak to peak for the defect-free structure to 0.19 \AA. Atomic displacements are thus small and range between 0.05 \AA\ and 0.13 \AA. Maximum displacements are observed for graphene atoms near the adatoms first neighbors of the defect. Conversely, adatoms of the reconstruction have rose toward graphene, with a maximum displacement of 0.11 \AA\ for the 3 adatoms neighboring the defect (for the C1 configuration: the adatoms labeled 1, 3 and 4 on Fig. \ref{fig2} (a)). The 4th adatom exhibits a displacement of 0.06 to 0.08 \AA, depending on the location of the defect.

Regarding the stability of the structures, the location of the defect does not seem to be a decisive factor. The total energy of C1, C3 and C4 differ only by maximum 2 meV and C2 differs from the 3 other configurations by maximum 9 meV per unit cell. The less stable configuration is C2. 

Thus, the presence of defects implies some structural modifications. The graphene-substrate distance tends to decrease, which could come with modifications of the electronic properties of the graphene layer and especially a strengthening of the graphene-substrate interaction, contrarily to what is observed on the Si face samples when H is introduced at the interface. We will address this question in the following.
\subsection{\label{elec_struct}Electronic structure}
\begin{figure}[!h]
\includegraphics[width=0.5\textwidth]{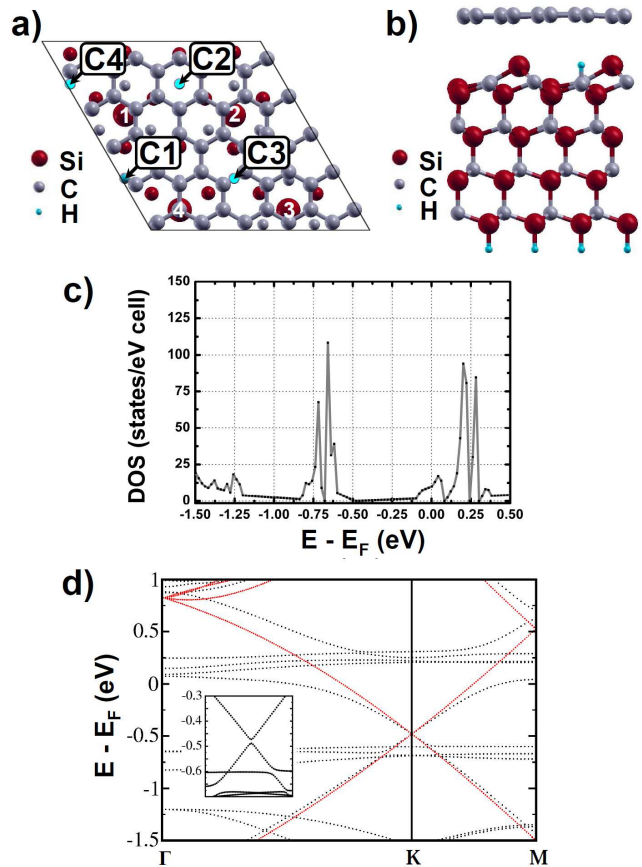}
 \caption{\label{fig2} (Color online) A $5\times 5$ graphene cell on top of a $4\times4$ $SiC(000\overline{1})$ substrate in the $(2\times2)_C$ reconstruction with one defect (H atom) per cell. (a) Top view of the supercell. Only the last SiC bilayer, the Si adatoms, the H atom and the graphene atoms are represented. Within the supercell (black diamond cell), all atoms are plotted once and their size refers to their height. The Si adatoms are numbered from 1 to 4 and the C1 to C4 labels refer to the 4 possible locations of the H atom. (b) Side view of the supercell after relaxation for the C1 configuration. (c) density of states for the C1 configuration. (d) Band structure for the C1 configuration in dotted line. The dispersion of an isolated graphene layer, shifted to align $E_D$s, is given in solid line. The inset shows a zoom of the C1 configuration band structure in the vicinity of the K point.}
\end{figure}

In the previous study\cite{magaudprb}, for the bare SiC$(2\times2)_C$ surface in the $4\times4$ geometry, the integrated DOS showed two narrow sets of bands corresponding to the adatom and restatom states, with the Fermi level at the top of the restatom band. Adding a graphene layer on top of the reconstruction did not qualitatively change the DOS of the interface (see the DOS of the defect free G$\_2\times2$ structure in Ref. \onlinecite{EPAPS}, Fig. S1 (b)).
When a defect is added to the structure - regardless of its location -, the total DOS shows a marked shift of the bands toward filled states compared to the DOS of the bare SiC$(2\times2)_C$ or the G$\_2\times2$(see Fig. \ref{fig2} (c) for the C1 configuration and Ref. \onlinecite{EPAPS}, Fig. S1 (a) for the other configurations). 

A detailed observation of the band structure provides more information on this aspect. The band structure of the C1 system is represented in Fig. \ref{fig2} (d). First of all, the $\pi$ bands of graphene are still present. The Dirac point is located approximately 0.48 eV below the Fermi level, indicating a significant electron doping. Bands are linear over $+/-100$ meV with respect to the Dirac energy ($E_D$). For higher energies with respect to $E_D$, a band anticrossing effect appears, involving the $\pi^*$ band and an adatom band, which distorts the graphene band around $E_F$. Thus, a hybridization of graphene states with adatom DBs occurs. This feature was already present in the defect free system but here, the energy splitting at anticrossing increases from $\Delta E \approx 0.35$ eV (defect-free structure) to $\Delta E \approx 0.45$ eV (see Ref. \onlinecite{EPAPS}, Fig. S2(a)) confirming an increased interaction with graphene. Regarding the $\pi$ band, interaction with states from the C restatoms remains very weak, the restatoms still being located far from the graphene layer. Finally, very small or vanishing gaps are observed at the Dirac point. C1 shows a gap of  $\approx 15$ meV (see Fig. 2 (d)) and the maximum gap of $\approx 30$ meV is obtained for C2 (see Ref. \onlinecite{EPAPS}, Fig. S2(b)).

\subsection{\label{comp_STM}Partial charge distribution versus STM images}
In order to check if our defect model accurately describes the real system, it is interesting to compare partial charge density maps to STM data. We however stress that such a comparison is limited to qualitative considerations because of the following reasons.
The real doping is certainly smaller than the one we get from ab initio calculations, due to a smaller density of defects in the real system. 
Thus, the relative positions of $E_D$ and $E_F$ are not quantitatively reproduced. Moreover, it is known that the DFT gap is systematically underestimated and the position of the dangling bond states bands is therefore only qualitative. Furthermore, the extension in vacuum of the numerical wavefunction is significantly underestimated. Thus, the low-bias STM images are compared to cross sections taken \itshape just above \upshape the graphene atoms. For the same reason, high-bias STM images - which reveal the SiC surface states - are compared to cross sections taken just above the adatoms, i.e. below the graphene layer.
Note finally that, due to the limited size of the supercell considered here, the calculation essentially provides information on the vicinity of the defect. Away from the defect, the STM image is reproduced by the defect-free interface model\cite{magaudprb}.

Fig. 3 (a) is a recall of the calculated structure (translated with respect to Fig. 2 (a) to get the defect in the middle of the cell). In the first instance, we concentrate on the states belonging to the substrate, which are probed on high-bias STM images. Thus, partial charge density maps are taken just above the adatoms (see the right part of Fig. 3 (a)).
Fig. 3 (b) is a cross section of $|\Psi|^2$ above the adatoms, integrated from $-1.0$ eV to $0.0$ eV. It has to be compared to high- bias filled-state STM images like in Fig. \ref{fig1} (b). As expected for the SiC($2\times2)_C$ reconstruction, states are present on the restatoms. However, at variance with the ideal reconstruction, a restatom is ``missing''(indicated by the arrow). It corresponds to the one bonded to the H atom, whose DB has been saturated by the H atom. Moreover, states are also detected on the adatoms, with higher intensity -even brighter than on the restatoms- on the 3 neighbors of the defect, labeled 1, 3 and 4 on Fig. 3. This higher intensity is in good agreement with the STM data in the vicinity of the defect site, except that calculations show similar charge density on the 3 neighbor adatoms while STM images usually do not. We will comment on this later. 

Fig. 3 (c) also shows a cross section above the adatoms, but for $|\Psi|^2$ integrated from $0.0$ eV to $+0.5$ eV, which has to be compared to high-bias empty-state images like in Fig. \ref{fig1} (c). In this case, states are localized on the adatoms, with equivalent intensity. In the empty-state STM image of Fig. \ref{fig1} (c), we also have signal only on adatoms, but with increased intensity on the neighbors of the defect. We however observed on positive sample bias STM images that the contrast near defects depends on the probed energy window. The larger the energy window, the less noticeable is the perturbation from the defect.
\begin{figure}
\includegraphics[width=0.4\textwidth]{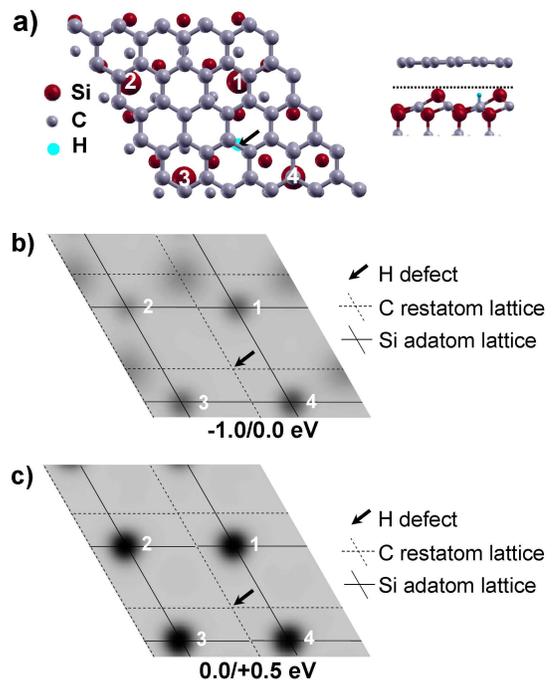}
 \caption{\label{fig3} (Color online) Partial charge density maps for the C1 configuration displaying the substrate states (Contrast is reversed with respect to Fig. 1: higher density is darker). (a) (left) Calculated C1 structure. The cell is translated with respect to Fig. 2 (a) so that the H atom, highlighted by an arrow, is in the middle of the image. The size of the atoms refers to their height in the supercell. The Si adatoms are numbered from 1 to 4, adatoms 1, 3 and 4 being the nearest neighbors of the defect. (Right) Side-view of the calculated structure, the dotted line indicates the section plane of the partial charge density maps. (b) and (c) Maps of $|\Psi|^2$ integrated from $-1.0$ eV to $0.0$ eV and from $0.0$ eV to $+0.5$ eV resp.}
\end{figure}

We now focus on the states belonging to the graphene layer. Thus, as illustrated in Fig. 4 (a), we consider partial charge density maps taken on a plane located just above the graphene atoms (Fig. 4 (b) and (c)), and also above the adatoms when the graphene-substrate interaction is considered (Fig 4 (d)). In the following, we will see how features characteristic of the defect free G\_$2\times2$ (see Fig. 1 (e), away from the defect) are reinforced due to an increase of the interaction, that is: supercell dark/bright halves, 3-fold symmetric line pattern and switched-off atoms\cite{magaudprb}.

Fig. 4 (b) shows a cross section of $|\Psi|^2$, just above the graphene layer, integrated from $-0.55$ eV to $-0.2$ eV. Notice that, at variance with Fig. 3, four supercells are represented. A schematic view of the data is given on the right part of the image, in which we also added the Si adatoms and the H defects as guides to the eye. As expected for the considered energy range, i.e. where the graphene/substrate interaction remains weak (energy window in between the SiC$(2\times2)$ dangling bonds states, Fig. 2 (d)), we observe essentially a honeycomb contrast. Within the supercell, which is indicated by a black diamond cell on Fig. 4 (b), the upper left half of the supercell appears higher (darker with this color scale) than the lower right one.  This feature arises from the topographic modulation described earlier in this paper. If we then consider a larger area than the supercell, we notice that the higher graphene atoms form a 3-fold symmetric pattern made of lines of hexagons (see the schematic view on the right part of Fig. 4 (b)).

In Fig. 4 (c) and (d), cross sections of $|\Psi|^2$ integrated from $-0.1$ eV to $+0.1$ eV are represented. In this energy range, the coupling between the electronic states of the graphene layer and the substrate states, namely the adatoms states, is strong (discussed in section III. B,  Fig. \ref{fig2} (d)). The left part of Fig. 4 (c) shows data taken just above the graphene layer, illustrated by a schematic view on the right part. The supercell is again indicated by a black diamond cell. In the schematic view, Si adatoms and H defects are added to the data. The graphene atoms within the supercell are represented by small black balls when they display significant charge density and by dotted circles when the charge density is much smaller. In the lower part of the 5x5 graphene supercell, we clearly see a contrast asymmetry between the graphene A and B sublattices. It seems to arise from the symmetry of the local stacking. Indeed, in this region, 3 graphene atoms belonging to the same graphene sublattice are located near adatoms and are switched off. In the upper part of the cell, one graphene atom, belonging to the other graphene sublattice and located on top of an adatom is also switched off.

If we then move down to the top of the adatom with Fig. 4 (d), charge density appears on the adatoms, mainly on the 3 neighbors of the defect. Actually, Fig. 4 (c) and (d) give a real space image of the interaction between graphene and the reconstruction adatoms. Fig. 4 (d) shows that hybridization occurs mainly between the $\pi$* band of graphene and the DBs of the 3 adatoms neighboring the defect. Finally, notice that high-bias empty-state images like in Fig. \ref{fig1} (c) in fact correspond to an average of data in Fig. 3 (c) and 4 (d), with a DOS signal on the Si adatoms, which is increased on the 3 neighbors of the defect. This illustrates the fact, evoked in the discussion in connection with Fig. 3 (c) that the aspect of the adatoms depends on the integration window.

\begin{figure}
\includegraphics[width=0.4\textwidth]{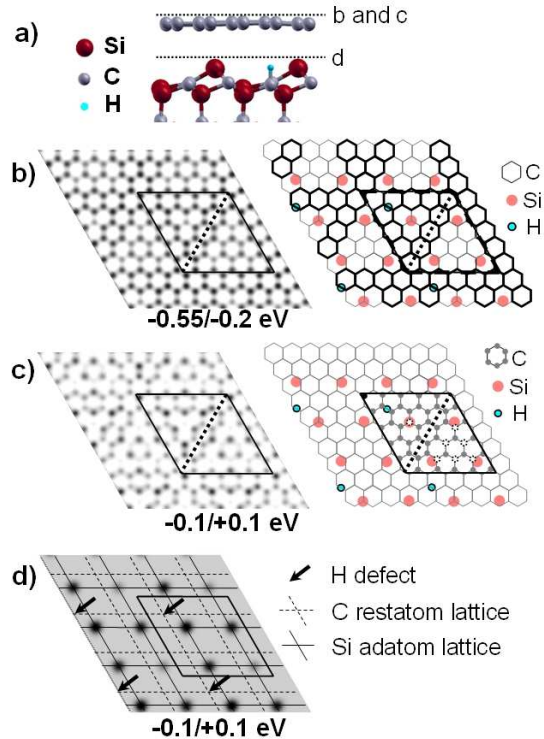}
 \caption{\label{fig3} (Color online) Partial charge density maps for the C1 configuration showing the substrate states (Contrast is reversed with respect to Fig. 1: higher density is darker). States from the graphene layer are considered here. (a) Side-view of the calculated structure, the dotted-lines indicate the height of the resp. partial charge density maps showed in the following. (b) (left) Map of $|\Psi|^2$ integrated from $-0.55$ eV to $-0.2$ eV (low-interaction region), taken just above the graphene atoms. The black diamond cell indicates the $5\times5$ graphene supercell, thus, 4 of them are represented on the image. (right) Corresponding schematic view. The honeycomb lattice represents the graphene lattice, higher density is darker. The Si adatoms and the H defects are also represented. (c) and (d) Maps of $|\Psi|^2$ integrated from $-0.1$ eV to $+0.1$ eV (interacting region) taken just above the graphene atoms and just above the adatoms resp. In (c) the data (left) is also completed by a schematic view (right). Within the supercell, the C atoms belonging to the graphene plane are depicted as small black balls and as dotted circles (see text). Again, Si adatoms and H defects are represented. In (d), the position of the H defect is highlighted by an arrow.}
\end{figure}
\section{\label{discussion} Discussion and conclusion}
To sum up, for the kind of defect we study, a restatom DB is saturated by an H atom. As a result, the charge transfer from the 3 first neighbors adatoms toward the restatom is impossible, which leaves a delocalized electron on these 3 adatoms. This local perturbation leads to a doping in electrons of the graphene layer. The Dirac point is namely shifted by $\approx 0.5$ eV under the Fermi level, which would correspond to $n \approx 2\ .10^{13} cm^{-2}$ for ideal graphene. Such a doping is of the order of the electron density created by the defect($n \approx 7.6\ .10^{13} cm^{-2}$ i.e. one electron per defect). Modifications of the electronic structure come with a maximum reduction by 0.2 \AA\ of the graphene- adatom distance. The graphene-substrate interaction is slightly enhanced, and involves the 3 adatoms neighboring the defect. However, the linear dispersion is preserved for energies within +/- 100meV with respect to the Dirac point. 

Based on STM results, our model seems to reproduce well the perturbations induced by the defect to the SiC($2\times2$). Calculations however show defects with higher symmetry than in the real system since the 3 adatoms neighboring the defect show similar charge density(Fig.3 (b)). Another calculation with a shift in the relative position between graphene and SiC lattices, while leading to a qualitatively similar electronic structure, results in a less symmetric defect in the direct space (see Ref. \onlinecite{EPAPS}, Fig. S3). Moreover, we actually considered the most simple defect configuration involving an H adsorbate while in the real system other species might come into play. Additionally, the system studied exhibits a high density of defects ($7.6\ 10^{13} cm^{-2}$) due to the technical constraint that limits the size of the supercell.  Such a high defect density might impose symmetries in the system. 

Considering the graphene-substrate interaction, because of the high density of defects in the calculated structure, the doping and the enhancement of the interaction with the substrate are most probably overestimated with respect to the real system. Other groups have measured an intrinsic doping in electrons of $n\approx 10^{12} cm^{-2}$\cite{CB, emtsevprb}. From the typical STM image in Fig. \ref{fig1} (a), we find a defect density of $n\approx 10^{13} cm^{-2}$. The experimental defect density is high enough to generate the experimentally measured electron concentration
and defects are thus a plausible cause for the doping of the graphene layer. 
Note that on STM low-bias images, no $(\sqrt{3}\times \sqrt{3})R30^\circ$ standing wave pattern are observed in the vicinity of defects. Thus, the defects -located at $\approx 3$\AA\ under the graphene layer- do not lead to intervalley scattering.

Finally, in our system, modification of the interface by introducing H atoms ($10^{14} cm^{-2}$) leads to a doping in electrons of the graphene layer and also slightly enhances the graphene-substrate interaction. Riedl et al. in Ref. \onlinecite{expeHintercale} have shown that, in the case of graphene grown on the Si face of SiC, the graphene-substrate interface can be H passivated and the graphene-substrate interaction vanishes. Starting from the strongly coupled buffer layer, after H treatment, they obtain a quasi-free-standing graphene layer, neutral, with linear dispersion. Thus, the molecular H migrates under the buffer layer and bonds to Si atoms of the substrate, breaking the covalent bonds between the buffer layer and the substrate\cite{DFTHintercale}. As a result, graphene characteristics are restored in the first graphitic layer. In the case of the C face, we conclude from our results that in order to decrease the graphene-substrate interaction, the density of H atoms introduced has to be sufficient to destroy the SiC($2\times2)_C$ reconstruction and saturate the ideal SiC surface. Otherwise the presence of H atoms would rather lead to a doped layer with an enhanced graphene-substrate interaction.  

\begin{acknowledgments}
This work is supported by a computer grant at 
the ACI CIMENT (phynum project); the french ANR
(GRAPHSIC project ANR-07-BLAN-0161, NANOSIM\_GRAPHENE project ANR-09-NANO-016-01); a CIBLE, and a RTRA DISPOGRAPH project. FH held a doctoral support from R\'{e}gion Rh\^{o}ne-Alpes. Figures
are plotted using Xcrysden.
\end{acknowledgments}

\end{document}